\documentclass[10pt]{article}
\usepackage{amsmath}
\usepackage{amssymb}
\usepackage{graphicx}
\usepackage{cite}
\usepackage{color} 
\usepackage{subfigure}
\usepackage{url}

\usepackage[labelfont=bf,labelsep=period,justification=raggedright]{caption}
\makeatletter
\renewcommand{\@biblabel}[1]{\quad#1.}
\makeatother
\date{}

\begin{document}

\begin{flushleft}
{\Large
\textbf{Self-organization on social media: endo-exo bursts and baseline fluctuations}
}
\\
Mizuki Oka$^{1}$, 
Yasuhiro Hashimoto$^{2}$,
Takashi Ikegami$^{3}$,
\\
\bf{1} Department of Computer Science, University of Tsukuba, 1-1-1 Tennnodai, Tsukuba 305-8577 Japan
\\
\bf{2} Department of Human and Engineered Environmental Studies, Kashiwa, Chiba 277-8577 Japan
\\
\bf{3} Graduate School of Arts and Sciences, The University of Tokyo, Komaba, Tokyo 153-8902 Japan
\\
$\ast$ E-mail: mizuki@cs.tsukuba.ac.jp
\end{flushleft}

\section*{Abstract}
A salient dynamic property of social media is bursting behavior. In this paper, we study bursting behavior in terms of the temporal relation between a preceding baseline fluctuation and the successive burst response using a frequency time series of 3,000 keywords on Twitter. We found that there is a fluctuation threshold up to which the burst size increases as the fluctuation increases and that above the threshold, there appears a variety of burst sizes. We call this threshold the critical threshold. Investigating this threshold in relation to endogenous bursts and exogenous bursts based on peak ratio and burst size reveals that the bursts below this threshold are endogenously caused and above this threshold, exogenous bursts emerge. Analysis of the 3,000 keywords shows that all the nouns have both endogenous and exogenous origins of bursts and that each keyword has a critical threshold in the baseline fluctuation value to distinguish between the two. Having a threshold for an input value for activating the system implies that Twitter is an \textit{excitable medium}. These findings are useful for characterizing how excitable a keyword is on Twitter and could be used, for example, to predict the response to particular information on social media.

\section*{Introduction}
\begin{figure*}[!hb]
\centering
  \includegraphics[width=0.80\textwidth,bb=0 0 3089 1700]{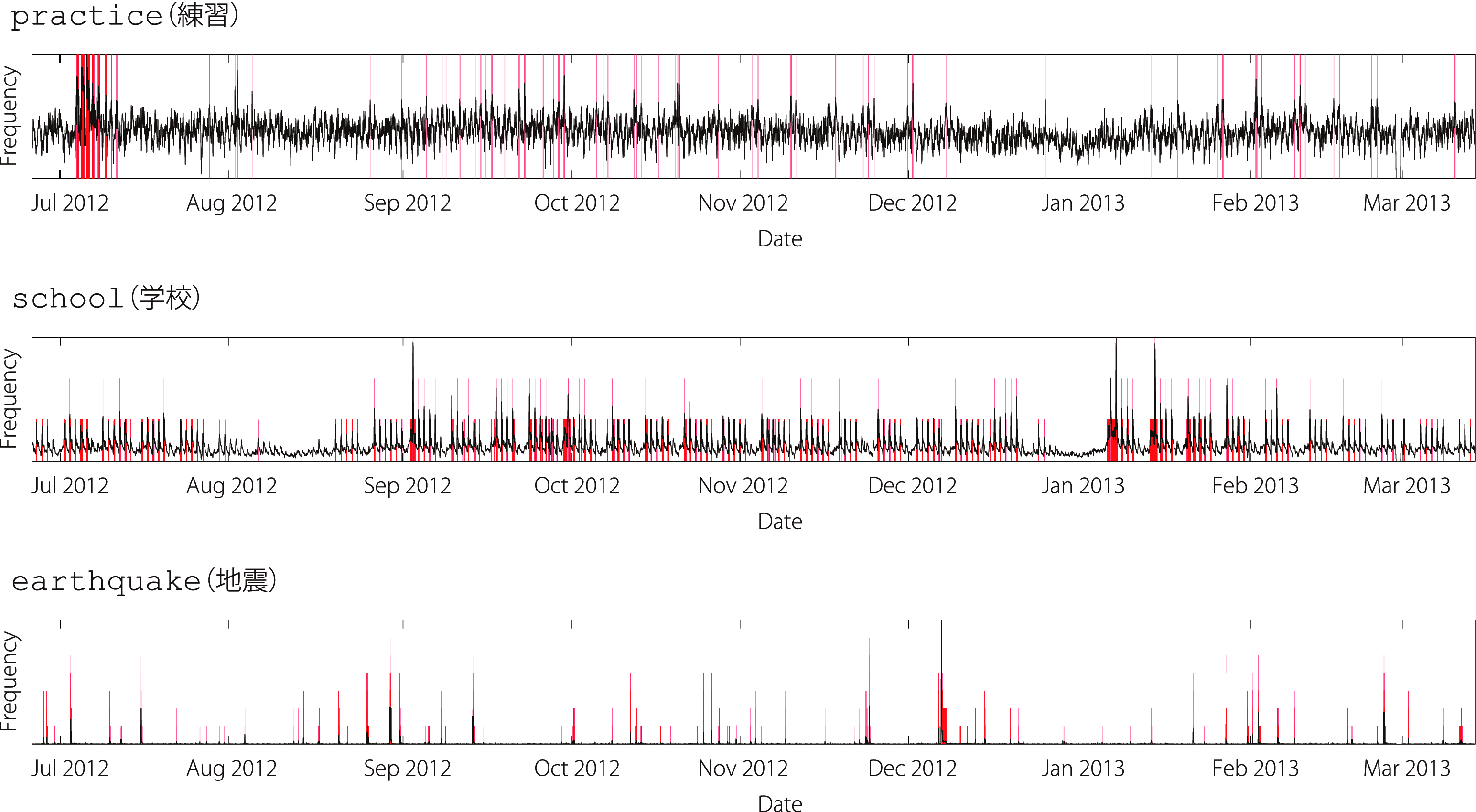}
\caption{ {\bf Examples of time series {\tt practice}(練習), {\tt school}(学校), and {\tt earthquake}(地震).} The original time series are depicted with black lines, and the detected bursts are depicted with red bars with the height indicating the burst level.
 \label{fig:fig1}}
\end{figure*}

Social media, such as Facebook, Twitter, and Google Plus, have established their role as information-sharing tools, both personally and commercially~\cite{Kwak10}. With the introduction of these new forms of social media, one can observe how people respond to specific information on the web. Indeed, social media have been widely used as platforms to study the emergence of patterns of collective attention~\cite{ferrara:traveling,lehmann2012,bollen2010,Kwak10}. When information receives collective attention, the information appears as a \textit{burst}, an increase in the number of appearances about the information for a certain period of time. For example, if we take the number of tweets that contain the keyword {\tt earthquake} as depicted in Figure~\ref{fig:fig1}, the bursts in the keyword time series show a strong correlation with the occurrences of earthquakes. This is because when there is an earthquake, people tend to tweet about it using the keyword {\tt earthquake}. These bursts occur aperiodically in accordance with the timing of earthquakes. Another example of bursts is observed as daily or weekly cycles in the keyword time series, such as {\tt school}, as depicted in Figure~\ref{fig:fig1}. We observe such daily periodic bursts because people attend school every day on weekdays and like to tweet about it. These kinds of periodic cycles have also been widely studied on social media and include individual-level diurnal and seasonal mood rhythms, which have been identified in cultures across the globe and are consistent with the effects of sleep and circadian rhythm~\cite{happiness_2011,diurnal_2011}. As these examples show, by aggregating the time series of keywords on social media, such as Twitter, we can extract patterns that exhibit underlying natural phenomena to human behavior. There is also a keyword, such as {\tt practice}, which shows only a few bursts but continuous fluctuations in the number of tweets, as shown in Figure~\ref{fig:fig1}.

Several studies have looked into the underlying mechanism that generates various bursting behaviors to reveal insights into people's collective behavior. For example, Conover et al. examined the temporal evolution of digital communication activity relating to the American anti-capitalist movement (i.e., Occupy Wall Street) using Twitter~\cite{occupy_wall_street2013}. The results indicated that the movement tended to elicit participation from a set of highly interconnected users with pre-existing interests in similar topics, such as domestic politics and foreign social movements. Some researchers also found early indicators on user-generated content of social media before large changes in events, such as movie box office success~\cite{moviebox_2013} or stock markets~\cite{google_trends2013,wikipedia_market2013}. Preis et al. found patterns that may be interpreted as early warning signs of stock market moves by analyzing changes in Google query volumes for search terms related to finance\cite{google_trends2013}. Similarly, Moat et al. investigated whether data generated through Internet usage contains traces of attempts to gather information before trading decisions are taken. They present evidence in line with the intriguing suggestion that data on changes in how often financially related Wikipedia pages were viewed may have contained early signs of stock market moves\cite{wikipedia_market2013}.

Other studies have looked more closely into the types of burst. For example, Crane and Sornette analyzed a property of a burst in terms of endogenous and exogenous bursts~\cite{crane2008}. Exogenous bursts are caused by external influences, such as earthquakes or appearances in the mass media. Endogenous bursts are caused as a result of word-of-mouth interactions in a social network. Crane and Sornette found that whether a burst is exogenous or endogenous can be found by looking at the peak ratio of the burst; when the peak ratio is \textit{small}, then the burst is endogenous, otherwise exogenous. Lehmann et al. applied their findings to a large-scale record of tweets, specifically hash-tagged tweets, and used endogenous and exogenous bursts to demonstrate that tweets can be clustered into four classes~\cite{lehmann2012}.

In this study, we are interested in the temporal as well as the internal structure that defines the origins of bursts. We first attempt to characterize a burst in relation to its temporal structure by investigating the relationship between a burst and fluctuation in the prior nonbursting period. We then investigate how the fluctuation plays a role in organizing endogenous or exogenous bursts. The fluctuation period is the period in which there is no outstanding increase in the keyword's popularity, which we refer to as the baseline period. For any keyword time series on Twitter, the baseline period is continuously fluctuating and burst sizes range from small to large. We hypothesize that as the baseline fluctuation increases, the burst size becomes larger. This kind of relationship is a generic application of fluctuation response in statistical physics\cite{reichl1998} in which a system's response size to an external stimulus has a linear relation with the size of the fluctuation. That is, the larger the fluctuation, the larger the response size. In general, the fluctuation-response relation holds in a thermal equilibrium system but is phenomenologically extended to many nonequilibrium open systems from physics~\cite{reichl1998} to biology~\cite{osawa1975,pnaskaneko2003} and economics~\cite{Ruelle2004}. We regard the size of burst as the strength of response on Twitter and the fluctuation in the number of occurrences of the keyword as the internal state of Twitter and show how endogenous and exogenous bursts are related to the level of fluctuation. Our findings reveal the emergence of various fluctuation-response relationships and the critical threshold in fluctuation size that divides endogenous and exogenous origins of burst.

\section*{Methods}

\subsection*{Data}
\begin{figure*}[!hb]
\centering
  \includegraphics[width=0.80\textwidth,bb=0 0 2617 700]{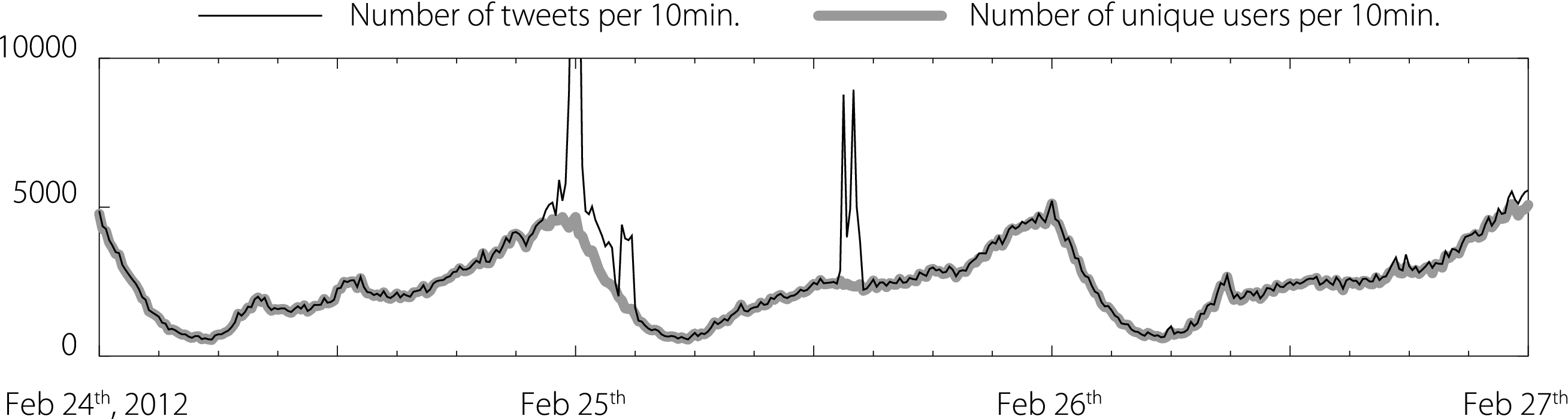}
\caption{ {\bf Number of tweets and unique users per 10-minute interval.} Sudden spikes in frequency indicate the impact of continuous posting of tweets by bots. The number of unique users mitigates the bot effect in the time series.
\label{fig:fig2}}
\end{figure*}

We collected tweets (in Japanese) over a 2-year period beginning July 2011, using the streaming APIs with the sampling method available for Twitter developers site~\footnote{Streaming API collects at most 1\% of all tweets produced on Twitter at a given time according to the documentation available at https://dev.twitter.com.}. We then applied morphological analysis using MeCab software, which is state-of-the-art software for Japanese morphological analysis\footnote{Available at https://code.google.com/p/mecab/.}. The collected data had many automated tweets posted by programs called bots, which resulted in peculiar statistics in the data. To mitigate the bot effect, we used the number of unique users to count the frequency of the keywords rather than the number of tweets (Figure~\ref{fig:fig2}). The basic statistics of the data are shown in Table~\ref{table:table1}. We chose the 3,000 most popular keywords from 1,550,770 distinct keywords and created a time series for each keyword by counting the number of unique users in 10-minute time intervals. We then smoothed each time series using a Gaussian kernel with a standard deviation of 30 minutes. This smoothing is applied to smooth out zero entries in the time series and ease handling of the data. The resulting time series is essentially equivalent to 1-hour time aggregation of the time series. We made the data available in terms of tweets IDs as well as each tweet IDs for each keyword of 3,000 keywords at http://dencity.jp/all\_ids.zip, and http://dencity.jp/3000.tgz, accordingly

\subsection*{Detection of bursts and fluctuations}
We symbolize a time series as a sequence of pairs of a baseline fluctuation period ($A_i$) and a following burst period ($B_i$). That is, a time series is translated into a sequence $A_1, B_1, A_2, B_2, \ldots, A_n, B_n$. Baseline and burst periods are determined by using the Kleinberg's burst detection algorithm\cite{kleinberg2002}.

The Kleinberg algorithm assumes the Poisson process for tweets; successive tweets occur independently following exponential distribution $f(x)=\bar{\lambda }e^{-\bar{\lambda} x}$, where $\bar{\lambda}$ is the overall mean frequency and $x$ is the interval of the successive tweets. $\bar{\lambda}$ is defined by $N/T$, where $N$ is the total number of tweets over the time series and $T$ is the total time length of the time series.

We calculate the ``burst level'' at each time $t$ in a time series, denoted as $i(t)$ (which takes integers). The burst level can be updated over time when the local mean frequency at time $t$, denoted as $\lambda_t$, exceeds a given threshold; if $\lambda_t$ exceeds $\bar{\lambda} s^1$, then the burst level $i(t)$ becomes $1$, and if it exceeds $\bar{\lambda} s^2$, then the burst level $i(t)$ becomes $2$, and so on. We set $s = 2$, so that the burst level increases by one when the frequency is twice as large as before. This way of changing the burst level may end up having a very large number of burst detections, including the noisy ones that have too short a duration. To mitigate this, another parameter $\gamma$ has been introduced in the algorithm. This controls the cost of changing the burst level between successive time points. In this study, we set $\gamma$ equals $1$. (For a more detailed explanation on the burst detection algorithm, see \cite{kleinberg2002}.) Given this setting, we define a ``burst period'' as the time period having the burst level larger than $0$ (i.e., $i(t) > 0$), otherwise, this is a ``baseline period''.

Using this algorithm, we labeled each period in a time series as either a baseline fluctuation period or a burst period for the 3,000 keyword time series. Figure~\ref{fig:fig1} shows examples of time series and their detected bursts for the keywords \textit{earthquake}, \textit{school}, and \textit{practice}. The original time series is depicted with black lines, and the detected bursts are depicted with red bars with the height indicating the burst level. We define the fluctuation by the standard deviation of the baseline frequency for each keyword $k$, denoted as
$$
\sigma_k(A_i) = \sqrt{\sum n_t^2 - (\sum n_t)^2},
$$ where $n_t$ denotes the frequency at time $t$ and $A_i$ is the $i$th baseline period in a time series. We also spotted a
time point when the frequency was the highest during the burst period;
we called this point the peak of the burst and denote it as
$P(B_i)$, where $B_i$ is the $i$th burst
period in a time series. We define the burst size, $S(B_i)$, as an integration of all
the frequencies in the burst period.

\subsection*{Classification of endogenous and exogenous bursts}
\label{sec:classification}
We identify each keyword's bursts as endogenous or exogenous by extending Crane and Sornette's work in~\cite{crane2008}. When the peak ratio becomes larger than a certain value, it is defined as exogenous bursts; otherwise, the burst is defined as endogenous. Crane and Sornette analyzed the property of a single burst; however, here we statistically analyzed a series of bursts and classified them as one of two distinct types. Namely, we considered not only the peak ratio but also the respective burst sizes to classify endogenous and exogenous bursts. More concretely, we measured each burst's peak-size ratio $P(B_i)/S(B_i)$ against its scaled burst size $S(B_i)/E(A_i)$.

Exogenous bursts form a pulse-like shape with the peak ratio $P(B_i)/S(B_i)$ becoming close to 1. Plotting all the exogenous bursts detected in a time series would result in a relatively constant high peak ratio, and they appear close to a constant value. Endogenous bursts do not form a pulse-like shape but a more flattened triangular shape. Assuming the peak of the burst to be relatively constant, then the peak ratio decreases as the burst increases. Thus, if we plot all the endogenous bursts in a time series, they appear closer to the $\alpha x^{-\beta}$ line. In sum, we can characterize endogenous and exogenous bursts as
\begin{itemize}
\item[1)] A small size burst tends to be the endogenous origin.
\item[2)] A large size burst tends to be the exogenous origin. 
\item[3)] A small peak ratio tends to be the endogenous origin. 
\item[4)] A large peak ratio tends to be the exogenous origin.
\end{itemize}
Namely, 1) and 3) above are endogenous origin bursts and 2) and 4) are exogenous origin bursts. On a two-dimensional plane, we find the exogenous points in the first quadrant and the endogenous ones in the third quadrant.

\begin{figure}[t]
\centering
\includegraphics[width=0.60\textwidth,bb=0 0 2958 2908]{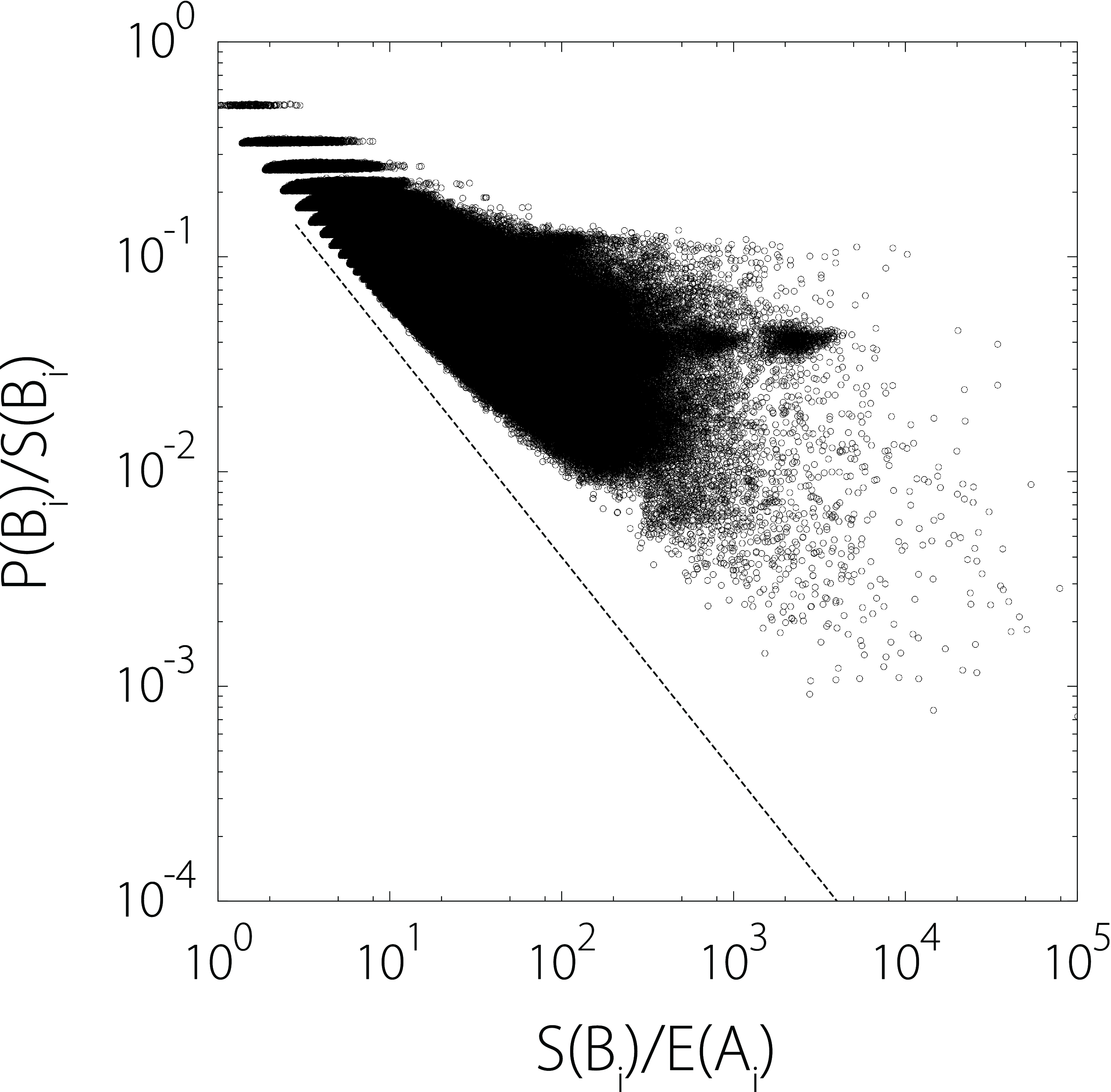}
\caption{ {\bf Overall burst size versus peak-size ratio for the 3,000 keywords.} Each keyword plot can be well-fitted by the power law distribution $\alpha x^{-\beta}$, where $(0< \beta <1)$ and $\beta$ is bounded at $1$.\label{fig:fig3}}
\end{figure}

The exogenous and endogenous bursts could be separated by the individually best fitted line as $y= \alpha x^{-\beta}$ and each keyword has a different ``separating line'' (or $\beta$.) However, if we separate points by the individually fitted line, we are implicitly separating bursts with almost equal ratio of endogenous and exogenous bursts. This is counterintuitive to our understanding that the endogenous and exogenous ratio should be different depending on the keyword. Therefore, we look for the common line of the slope, or $\beta$, to classify endogenous and exogenous bursts. So the question is how to set a parameter $\beta$ for all the time series to distinguish between them. For this, we plot the scaled burst size (x-axis) versus the peak-size ratio (y-axis) for all 3,000 keywords overlaying on top of each other, as shown in Figure~\ref{fig:fig3}. The 3,000 keywords are bounded between $x^{-1}$ and a constant value, and they can be well fitted by the power law distribution, which is $\alpha x^{-\beta}$ where $(0< \beta <1)$. Notice that the slope $-1$ is the lowest exponent out of the whole-fitted lines. This means that all the endogenous bursts are bounded at $\beta = 1$; thus, we set $\beta$ to be $1$ and use it as a global separating line. If a burst is below $y = \alpha x^{-1}$ (or a diagonal line $y = \alpha - x$ on a logarithmic scale), we regard the burst endogenous, otherwise exogenous. With Crane and Sornette's method, the peak ratio to distinguish between the two types of bursts is determined arbitrarily and is usually set empirically. Our proposed method does not require such a predefined threshold but rather is automatically set appropriately for each keyword while still respecting the peak-ratio criterion.

\subsection*{Representative Keywords}
\begin{figure*}[!hb]
\centering
 \includegraphics[width=1.00\textwidth, bb=0 0 4250 1494]{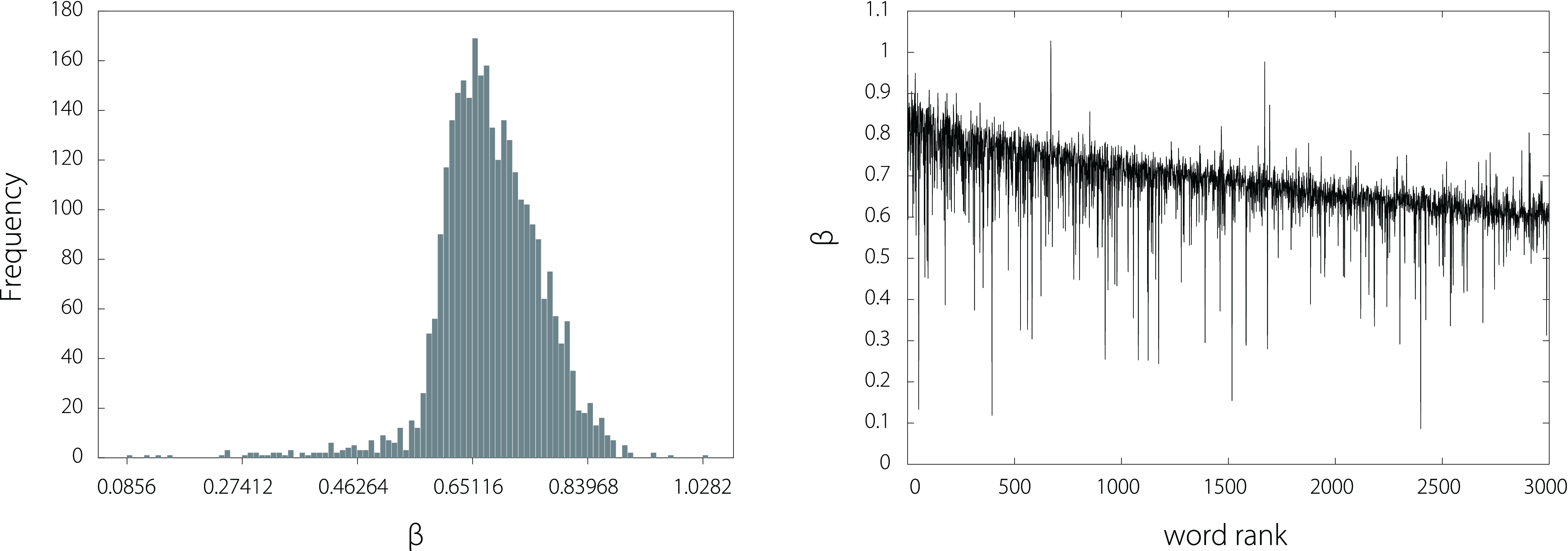}
\caption{{\bf Histogram of fitted $\beta$ for $\alpha \times x^{-\beta}$ of all 3,000 keywords (Left) and plot of $\beta$ with respect to word rank.}
\label{fig:fig4}}
\end{figure*}

Using this classification method, we automatically labeled all the bursts in each time series as endogenous or exogenous. If we plot all the fitted $\beta$ with respect to the keyword rank (ranked according to the total frequency), it shows a tendency that higher ranked keywords have larger $\beta$ values and lower ranked keywords have lower $\beta$, as shown in Figure~\ref{fig:fig4}. This indicates that higher ranked keywords tend to be more endogenous than exogenous and lower ranked keywords tend to be more exogenous than endogenous. Indeed, Table~\ref{table:table2} lists the keywords of the 10 highest $\beta$, which show daily frequently used keywords, such as {\tt food}, {\tt laugh}, {\tt person}, and so on. On the other hand, the 10 lowest $\beta$ show more externally driven keywords, such as {\tt seismic intensity} and {\tt earthquake}.

We show a few representative plots taking from different $\beta$ values. Figure~\ref{fig:fig5} shows three plots of the keywords {\tt practice}(練習), {\tt school}(学校), and {\tt earthquake}(地震) with $\beta = 0.85, 0.56, 0.12$, respectively, from more endogenous to more exogenous keywords. The endogenous bursts are shown in blue and exogenous bursts in red. As seen in these examples, most of the keywords have both endogenous and exogenous bursts, and the ratio of these types of bursts differs depending on the keywords.

\section*{Results}

\subsection*{Fluctuation-response relation and threshold}
We plot the temporal relation between a baseline fluctuation and a burst for each keyword by plotting each transition from $A_i$ to $B_i$ for $n$ number of pairs. As illustrative examples, Figure~\ref{fig:fig6} shows the plots for the keywords for {\tt practice}, {\tt school}, and {\tt earthquake}. They are also shown in blue or red, corresponding to endogenous and exogenous, respectively.

The keyword {\tt practice} shows a case where the response size (i.e., the maximum size, or the peak $P(B_i)$ of the burst period $B_i$) is correlated with the amplitude of the immediately preceding baseline fluctuation $\sigma(A_i)$. The keyword {\tt school} shows a case in which there is a point up to which the fluctuation gradually amplifies in correlation with the burst size and the fluctuation-burst relation then changes qualitatively and causes large bursts. We call this the critical threshold. Below this critical threshold, the burst response has a positive correlation with the preceding baseline fluctuation. Above the critical threshold, the size of the response becomes independent from the fluctuation size. Sometimes, these keywords have occasional bursts due to events that break periodicities. Taking the example of {\tt school}, some periodicities originate in the circadian rhythm. Sometimes this periodicity breaks and the fluctuation increases, causing the bursts that follow to also be larger (see Figure~\ref{fig:fig7}). These periodicity-broken phases correspond to major school breaks, such as spring, summer, and winter holidays. A disruption in repetitive everyday life triggers a large burst. The keyword {\tt earthquake} shows a case where the fluctuation-independent bursts range from small to large and merge at or above the critical threshold.

\subsection*{Endogenous and exogenous bursts}
Having classified the exogenous and endogenous bursts, we further investigate the relationship between the occurrence of bursts and the preceding baseline fluctuations. Namely, we examine the emergence of the critical threshold in the fluctuation-response relation in terms of endogenous and exogenous bursts. Our hypothesis is that the critical threshold emerges as a result of the two types of bursts so that they are separated at the threshold. That is, we consider a burst as endogenous if its corresponding baseline fluctuation is \textit{smaller} than the threshold. We consider a burst as exogenous if its corresponding baseline fluctuation is \textit{larger} than the threshold. We test this hypothesis by drawing a receiver operation characteristic (ROC) curve with a false positive rate on the x-axis and a true positive rate on the y-axis. A false positive is a misclassification of the burst, i.e., an endogenous burst detected as exogenous, whereas a true positive is a correct classification of the burst, i.e., an endogenous burst detected as endogenous. The false positive and true positive rates change according to the threshold. The ROC curve shows how well bursts can be classified correctly by changing the threshold. The closer the curve gets to the upper left corner, the better the classification is.

We computed an ROC curve for each keyword, counting the total 3,000 ROC curves, and plotted the result in Figure~\ref{fig:fig8}. Each ROC curve is plotted in gray, and the average curve is plotted in black. The area under the curve (AUC) can be computed to evaluate the how well the threshold can classify (The AUC becomes $0.5$ when random classification is done with ROC curve on the diagonal line). The AUC of the average ROC curve is $0.8545$ in our results. This is an indication that the critical threshold emerges as a result of endogenous and exogenous bursts and that this threshold can be used to separate the two. At the same time, the endogenous and exogenous causes are not always distinguishable. Exogenous bursts are followed by retweets, and endogenous bursts are implicitly affected by real world events. Henceforth, these two causes are intermingled, which is reflected in the continuous spectrum of the fluctuation-response curve, especially with many lower rank keywords. An analysis of ROC validates this hypothesis.

\subsection*{Burst size distributions}
\begin{figure*}[!hb]
\centering
  \includegraphics[width=1.00\textwidth,bb=0 0 3758 1192]{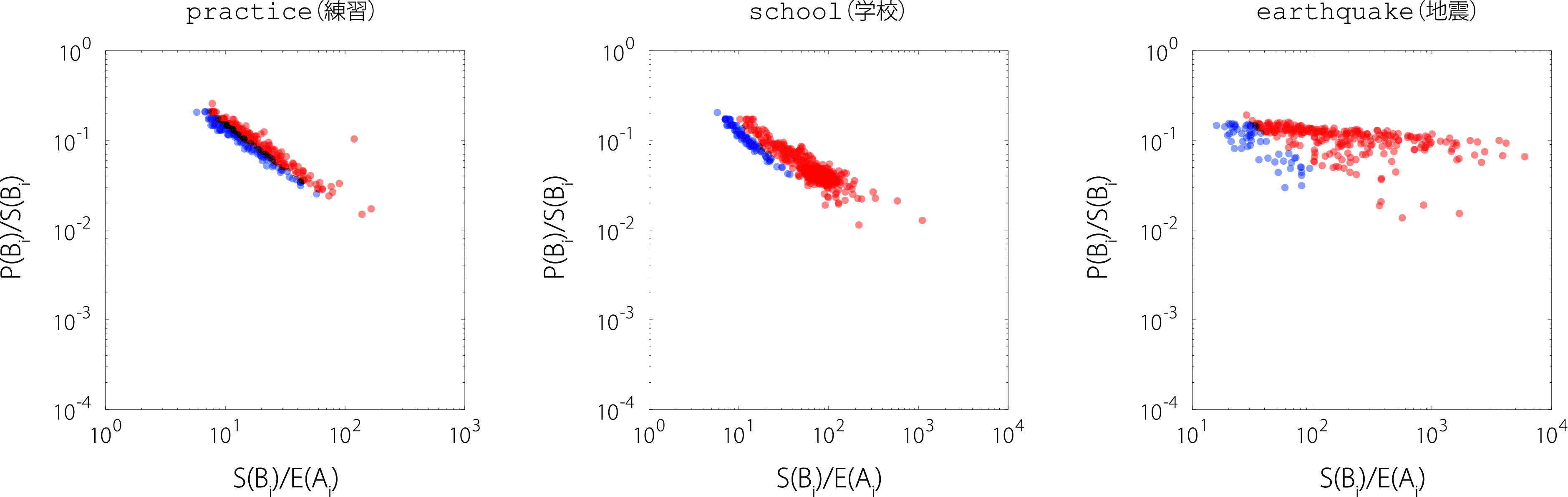}
\caption{ {\bf Burst size versus peak-size ratio.} The peak-size ratio is either inversely proportional to the burst size or becomes constant. $P(B_i)$ is the burst peak height, $S(B_i)$ is the burst size and $E(A_i)$ is the mean frequency in the baseline period. The size of the plots is proportional to $p_i$. The burst size versus the peak-size ratios for {\tt practice}(練習) (Left), {\tt school}(学校) (Middle), and {\tt earthquake}(地震) (Right). The exogenous (in red) and endogenous (in blue) are identified by fitting the points with the equation $y = \alpha - x$ (on a logarithmic scale) and labeling a point below the fitted line as endogenous and otherwise exogenous.
  \label{fig:fig5}}
\end{figure*}

\begin{figure*}[!hb]
\centering
  \includegraphics[width=1.00\textwidth,bb=0 0 3753 1208]{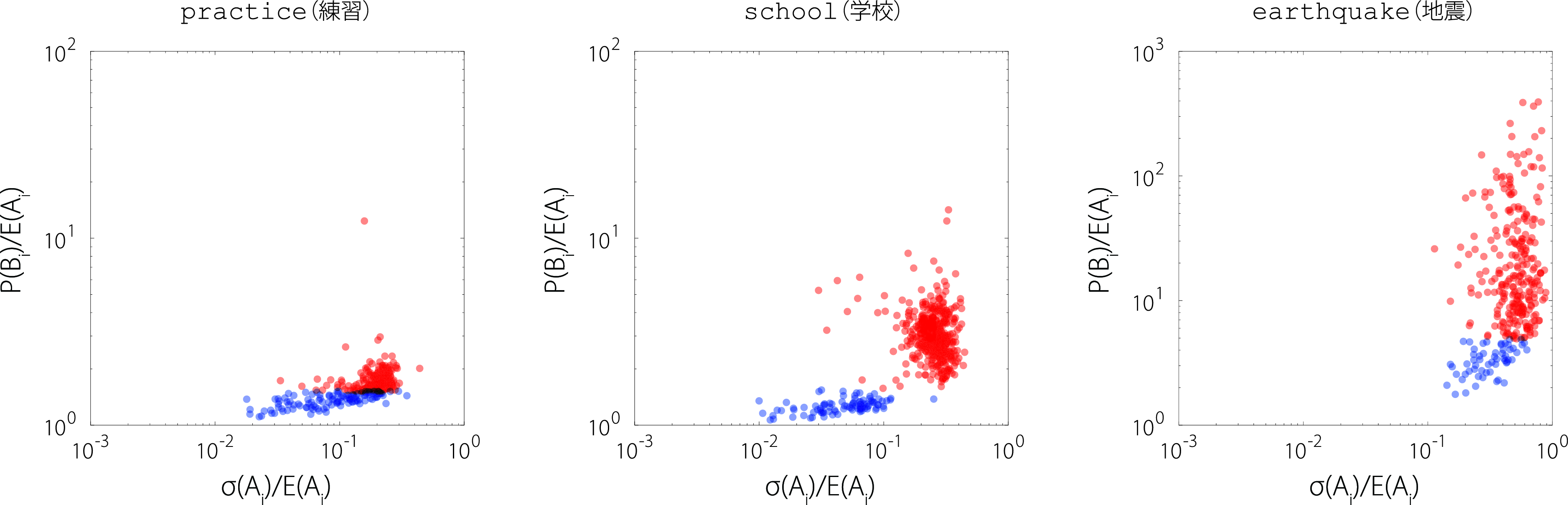}
\caption{{\bf Baseline fluctuation, $\sigma (A_i)/E(A_i)$ and peak burst size, $P(B_i)/E(A_i)$ for the keywords {\tt practice}(練習), {\tt school}(学校), and {\tt earthquake}(地震).} Points are in blue for endogenous and in red for exogenous according to the detection identified in the burst size versus peak-size ratio. Keyword {\tt practice}: the response size has a positive correlation to the amplitude of the baseline fluctuation that immediately precedes it. Keyword {\tt school}: has a positive correlation between the response size and the amplitude of the baseline fluctuation immediately preceding it to a certain threshold and has relatively large responses beyond the threshold. Keyword {\tt earthquake}: Abrupt responses ranging from small to large at a specific threshold; most importantly, all responses are concentrated around the threshold.
\label{fig:fig6}}
\end{figure*}

\begin{figure*}[!hb]
\centering
  \includegraphics[width=1.00\textwidth,bb=0 0 4058 625]{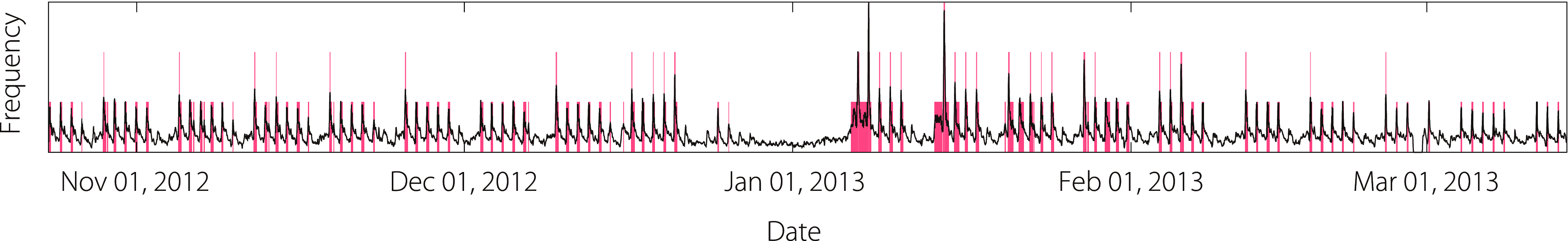}
\caption{ {\bf Breaks of the periodicities for the keyword school originating in the circadian rhythm causing the following bursts to also be larger.}
\label{fig:fig7}}
\end{figure*}

\begin{figure*}[!hb]
\centering
\includegraphics[width=0.60\textwidth, bb=0 0 733 728]{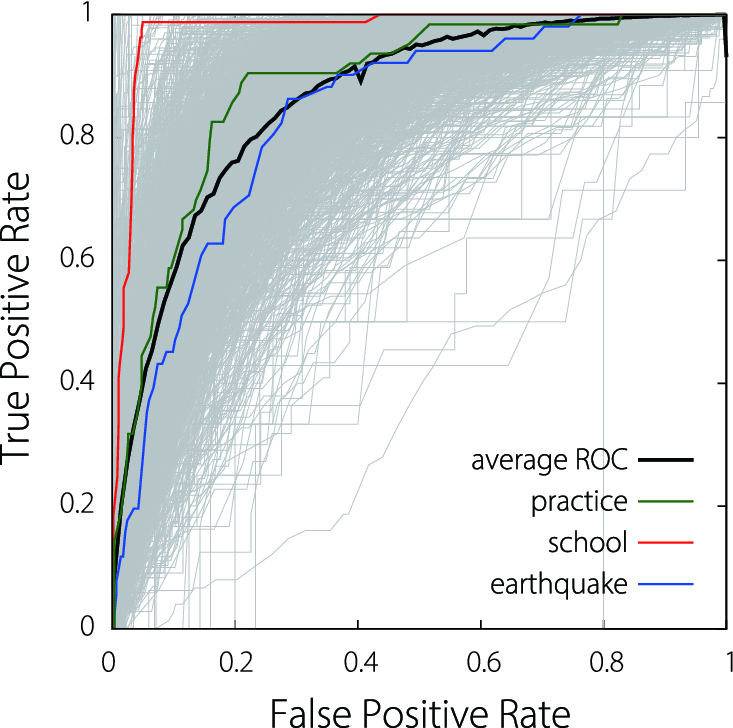}
\caption{ {\bf ROC curves for all 3,000 keywords.} Each ROC curve is plotted in gray, and the average curve is plotted in black. The AUC of the average ROC curve is $0.8545$. The ROC curves for {\tt practice}(練習), {\tt school}(学校), and {\tt earthquake}(地震) are shown in green, red and blue, respectively. The threshold is ranged for each keyword to compute false positive rates and true positive rates.
\label{fig:fig8}}
\end{figure*}

\begin{figure*}[!hb]
\centering
  \includegraphics[width=1.00\textwidth, bb=0 0 7217 2283]{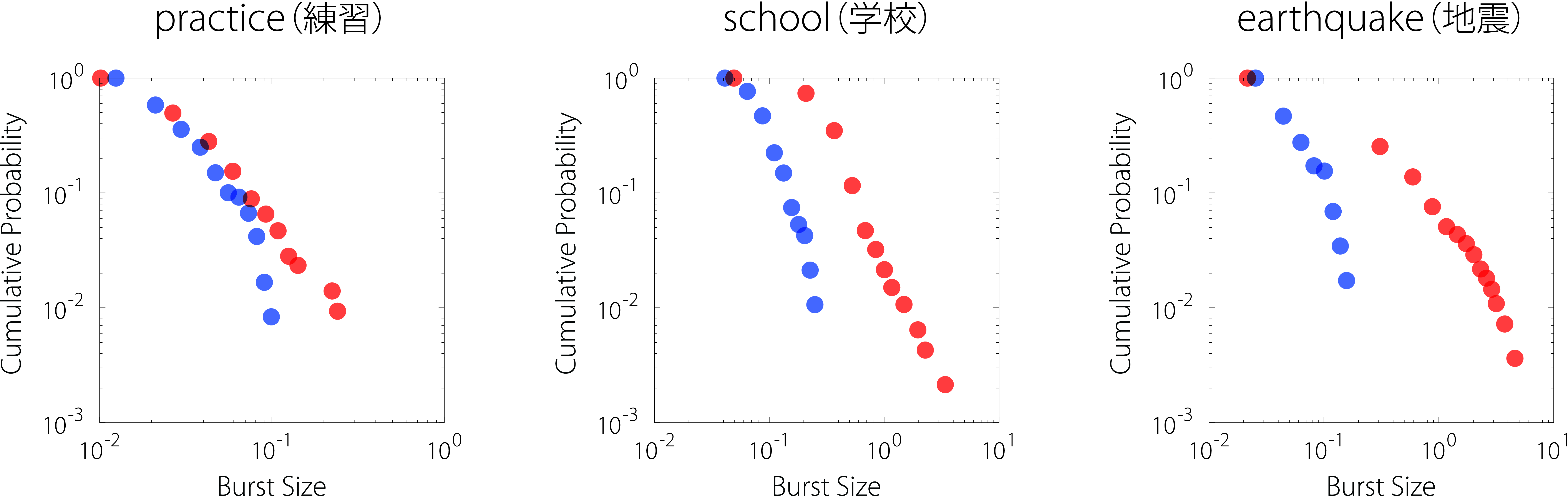}
\caption{ {\bf Cumulative size distributions of the endogenous and exogenous origins of bursts for the keywords {\tt practice}(練習), {\tt school}(学校), and {\tt earthquake}(地震).} The endogenous bursts mostly have smaller size bursts compared to the exogenous bursts. The gaps in the burst sizes between the two types increase from {\tt practice} to {\tt school} to {\tt earthquake}, which also indicates that the distinction between the two types becomes clearer. The distributions of exogenous bursts tend to show power law behaviors common to other human behavior statistics\cite{barabasi_2006,bak1987}.
\label{fig:fig9}}
\end{figure*}

Another way to look into the origins of bursts is to compute the burst size distribution. We examined the organization of the distribution to study the origins of thresholds. Figure~\ref{fig:fig9} shows the cumulative size distributions of bursts for the keywords {\tt practice} (練習), {\tt school} (学校), and {\tt earthquake} (地震). The figure plots the distributions using the automatically classified endogenous and exogenous origins of bursts as in Figure~\ref{fig:fig5}. For all the distributions shown here, the endogenous bursts mostly have smaller sized bursts compared to the exogenous bursts. Together with the findings shown in Figure~\ref{fig:fig5}, we can say that exogenous bursts have large-sized pulse-like bursts and endogenous bursts have small-sized flattened bursts. The gap in the burst sizes between the two types becomes larger from {\tt practice} to {\tt school} to {\tt earthquake}. This also means that when $\beta$ (the fitted value to $\alpha x^{-\beta}$ described in the section Classification of endogenous and exogenous bursts) is smaller (e.g., {\tt earthquake}), then the distinction between endogenous and exogenous bursts is clearer; however, when it is larger (e.g., {\tt practice}), then this distinction becomes more arbitrary. Indeed, we say that the bursts in {\tt practice} are mostly the endogenous ones since all the endogenous bursts are bounded at $\beta = 1$ and the $\beta$ of {\tt practice} is close to 1. 

\begin{figure*}[!hb]
\centering \includegraphics[width=0.60\textwidth,bb=0 0 1146 1083]{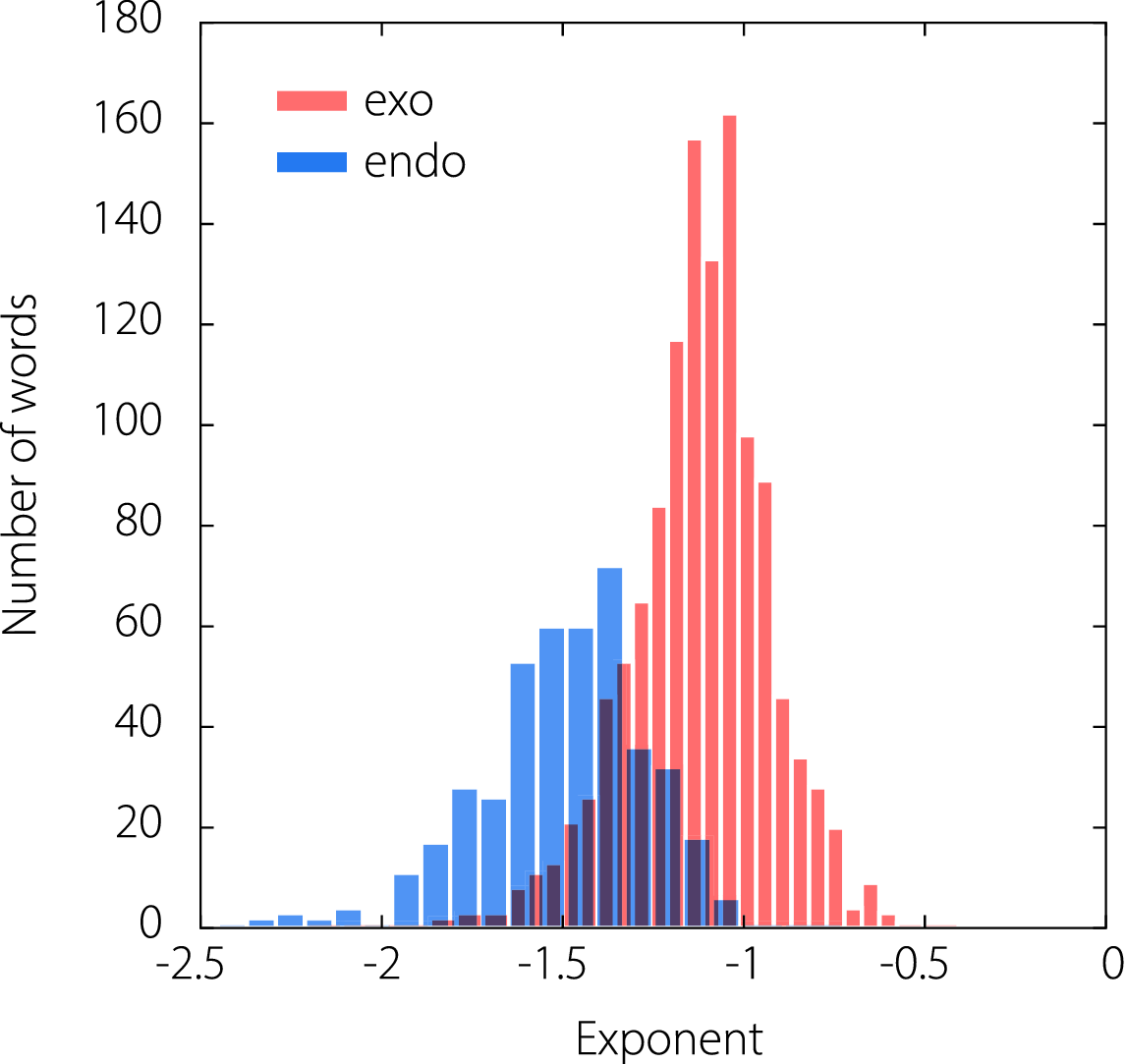}
\caption{ {\bf Histogram of exponents that satisfy the coefficient of determination $R^2>0.96$ of endogenous and exogenous bursts fitted to the power law when removing keywords whose distributions have less than 5 points.} The 434 and 1,240 keywords (out of 3,000) satisfy $R^2>0.96$ for endogenous bursts and exogenous bursts, respectively.
\label{fig:fig10}}
\end{figure*}

It is interesting to note that the distributions of bursts tend to show power law behaviors. When we fit both the endogenous and exogenous distributions with the power law distribution, the total of 434 of the endogenous burst's distribution and 1,240 of the exogenous burst's distribution satisfy the coefficient of determination $R^2>0.96$ \footnote{We removed keywords that have less than 5 points in the distributions.}. This shows that exogenous bursts tend to show power law behaviors more than endogenous bursts. The histograms of the fitted exponents for endogenous bursts (in blue) and exogenous bursts (in red) are shown in Figure~\ref{fig:fig10}. Since the exponents are cumulative distributions, the bare exponents are obtained by adding 1 to them. In Figure~\ref{fig:fig10}, we notice that all the exponents of the endogenous bursts are less than $-1$, so that the expected value of the burst size is bounded as their bare exponents are less than $-2$. Whereas half of the exogenous bursts have their exponents larger than $-1$, so that the expected value of burst size diverges. This tells us that the exogenous bursts size is not predictable and is consistent with our fluctuation-response relation described above; the various sizes of exogenous bursts start to emerge at and above a critical threshold. 

The power law behaviors are common to other human behavior statistics \cite{barabasi_2006, bak1987}. This means that the nature of bursts is scale free. A mechanism of organizing a single burst has shown that the underlying mechanism is modeled with a simple epidemic propagation model\cite{sornette2004}. If Twitter can be approximated by an epidemic growth model, the size distribution of bursts would follow a power law behavior, as suggested in \cite{bak1987}, known as a self-organized critical state. Or it can be explained as common of human ``queuing'' behavior\cite{barabasi_2006}. A detailed analysis of the scale-free nature in Twitter bursts remains as a future study.

\section*{Discussion}
Our findings suggest that the fluctuation threshold separates two natures of endogenous and exogenous bursts, but the classification based on the peak-ratio and burst-size plot as shown here is not perfect. Some bursts appear as a mix of endogenous and exogenous bursts, and they are not separable by a single fluctuation value. This is reflected in the ROC curve analysis. Moreover, when most of the bursts are around a curve of the exponent $-1$ on the peak-ratio and burst-size plot, all of them may correspond to endogenous bursts. Nonetheless, in our classification they are classified into exogenous and endogenous by the exponent of $-1$, and this remains as a limitation of our method. The mix of the two types of bursts may be reflecting essential human behaviors. 

It should also be noted that the observation of the fluctuation response (or burst) relationship does not imply that any fluctuation actually ``causes'' bursts. Our results show that most of the large bursts happen only at or beyond a critical fluctuation value and that they are mostly exogenous bursts. This implies that when the baseline fluctuation is larger, the system can amplify external influences into larger bursts. On the other hand, when the baseline fluctuation is smaller, the size of the bursts is relative to the fluctuation size and they are mostly endogenous bursts.

As an interpretation of the present results, we argue that a fluctuation and burst relationship reflects a shared feeling on Twitter when people become sensitive to certain information. Probably a shift of threshold has occurred when people become sensitive to a topic. A popular topic may lower the threshold and get it ready for reacting with a subtle trigger. In the previous study of the Twitter time series, we demonstrated that everyday keywords (the higher ranked keywords) tend to become information sources to the lower ranked words, which we named \textit{default mode states} of Twitter. Our findings here enforce the view of the default mode state in Twitter~\cite{oka2013}. 

\section*{Conclusions}
We studied endogenous and exogenous origins of bursts and their temporal relationship between a baseline fluctuation and the subsequent response as a burst. Results suggest that Twitter does not simply have a sensor that responds to stimuli from the outside but it also has a sensor that responds to internal dynamics. Responses to external stimuli emerge as exogenous bursts, while those to internal dynamics emerge as endogenous bursts. Almost all the keywords exhibit both exogenous and endogenous bursts, and the difference in the response can be characterized by the relationship between the baseline fluctuation and the burst sizes. An endogenous response increases along with a baseline fluctuation, while an exogenous response does not show such a correlation but shows unpredictable behavior irrespective of a baseline fluctuation. A fluctuation threshold that separates these two types of bursts emerges as the critical threshold. At or above the threshold, the response becomes unpredictable, showing a wide range of burst sizes as a result of external influences. The threshold has different values for different keywords and is self-organized, indicating that the different keywords have different sensitivity to the corresponding impact in the real or virtual world. Possible applications based on these findings are numerous. One can use Twitter, for example, as a sensor system for predicting future bursts of each keyword.

\bibliography{biblio}

\section*{Figure Legends}

\begin{table*}[!hb]
\centering
\caption{General statistics for the dataset
\label{table:datastat}}
\begin{tabular}{lr}
\hline
total number of tweets & 297,792,366 \cr
total number of users & 12,677,098\cr
total number of keywords & 1,550,770 \cr
  - used by more than two users & 623,218 \cr
  - used by more than 10 users & 276,867 \cr
total number of tweets in the top 3000 keywords & 162,358,768 \cr
total number of users in the top 3000 keywords & 12,037,771 \cr
\hline
\end{tabular}
\label{table:table1}
\end{table*}

\begin{table*}[!hb]
\centering
\caption{The top 10 largest and smallest $\beta$ keywords.
\label{table:table2}}
\begin{tabular}{l || l }
\hline
Top 10 largest $\beta$ & Top 10 smallest $\beta$\cr
\hline
food (食) & seismic intensity (震度)\cr
adult (成人)    & earthquake (地震)\cr
real (ほんと)    & morning (朝)\cr
laugh (笑)    & Pretty Cure (プリキュア)\cr
gold (金) & luck (運)\cr
thing (もの)   & health (健康)\cr
friend (友達) &  money luck (金運)\cr
day (日)      & fortune (運勢)\cr    
phone (phone)  & snack (おやつ)\cr
line (line)     & Ibaraki (茨城) \cr
\hline
\end{tabular}
\end{table*}

\end{document}